%% file: ngc300_ulx1.tex
\documentclass[twocolumn]{aastex63}
\usepackage{amsmath}
\usepackage{booktabs}
\usepackage{xspace}
\usepackage{enumitem}
\usepackage{hyperref}
\usepackage{chronosys}

\newcommand{\red}[1]{\textcolor{red}{#1}}

\accepted{October 10, 2022}
\submitjournal{ApJ}

\shorttitle{NGC300 ULX-1 Spectroscopy}
\shortauthors{Ng et al.}

\newcommand{\nicer}{NICER\xspace}
\newcommand{\xmm}{XMM-Newton\xspace}
\newcommand{\chandra}{Chandra\xspace}
\newcommand{\nustar}{NuSTAR\xspace}

\newcommand{\swift}{Swift\xspace}

\graphicspath{{./}{figures/}}

\begin{document}

\title{Spectral Evolution of Ultraluminous X-ray Pulsar NGC 300 ULX-1}

\correspondingauthor{Mason Ng}
\email{masonng@mit.edu}

\author[0000-0002-0940-6563]{Mason Ng}
\affiliation{MIT Kavli Institute for Astrophysics and Space Research, Massachusetts Institute of Technology, Cambridge, MA 02139, USA}

\author[0000-0003-4815-0481]{Ronald A. Remillard}
\affiliation{MIT Kavli Institute for Astrophysics and Space Research, Massachusetts Institute of Technology, Cambridge, MA 02139, USA}

\author[0000-0002-5872-6061]{James F. Steiner}
\affiliation{Smithsonian Astrophysical Observatory, 60 Garden St., Cambridge, MA 02138, USA}

\author[0000-0001-8804-8946]{Deepto Chakrabarty}
\affiliation{MIT Kavli Institute for Astrophysics and Space Research, Massachusetts Institute of Technology, Cambridge, MA 02139, USA}

\author[0000-0003-1386-7861]{Dheeraj R. Pasham}
\affiliation{MIT Kavli Institute for Astrophysics and Space Research, Massachusetts Institute of Technology, Cambridge, MA 02139, USA}

\begin{abstract}

We report on results from a one-year soft X-ray observing campaign of the ultraluminous X-ray pulsar NGC 300 ULX-1 by the Neutron star Interior Composition Explorer (NICER) during 2018--2019. Our analysis also made use of data from Swift/XRT and XMM-Newton in order to model and remove contamination from the nearby eclipsing X-ray binary NGC 300 X-1. We constructed and fitted a series of 5-day averaged NICER spectra of NGC 300 ULX-1 in the 0.4--4.0 keV range to evaluate the long-term spectral evolution of the source, and found that an absorbed power-law model provided the best fit overall. Over the course of our observations, the source flux (0.4--4.0 keV; absorbed) dimmed from $2\times10^{-12}$ to below $10^{-13}{\rm\,erg\,s^{-1}\,cm^{-2}}$ and the spectrum softened, with the photon index going from $\Gamma\approx1.6$ to $\Gamma\approx2.6$. We interpret the spectral softening as reprocessed emission from the accretion disk edge coming into view while the pulsar was obscured by the possibly precessing disk. Some spectral fits were significantly improved by the inclusion of a disk blackbody component, and we surmise that this could be due to the pulsar emerging in between obscuration episodes by partial covering absorbers. We posit that we observed a low-flux state of the system (due to line-of-sight absorption) punctuated by the occasional appearance of the pulsar, indicating short-term source variability nested in longer-term accretion disk precession timescales.

%We provided qualitative confirmation of the idea of a possibly precessing accretion disk or outflows obscuring the pulsar along the line-of-sight.

%The analysis was complicated by the nearby source NGC 300 X-1, a 32.8 hr eclipsing Wolf-Rayet black hole X-ray binary within the same NICER field-of-view. We made use of archival Swift/XRT and XMM-Newton/EPIC data to characterize the spectral emission for the ``on-eclipse" and ``off-eclipse" phases of NGC 300 X-1. The resulting spectra (of NGC 300 X-1) were combined with the modeled X-ray particle background to create a `total background' that were subtracted from the NICER spectra to isolate the NGC 300 ULX-1 emission. 

\end{abstract}


\section{Introduction} \label{sec:intro}

Ultraluminous X-ray sources (ULXs) are extremely bright, off-nuclear objects that have luminosities (assuming isotropic emission) of at least $L_X > 10^{39}{\rm\,erg\,s^{-1}}$ \citep[see][for a review]{kaaret17}, exceeding the Eddington limit of a 1.4 $M_\odot$ neutron star ($L_X \approx 2\times10^{38}{\rm\,erg\,s^{-1}}$), and are believed to be powered by super-Eddington accretion onto compact objects \citep{gladstone09,sutton13,bachetti14,fabrika21,qiu21}. Hundreds of ULXs and over a thousand ULX candidates have been discovered thus far \citep{pasham14,pasham15,earnshaw19,kovlakas20,walton22}, and they were originally thought to be entirely due to sub-Eddington accreting intermediate-mass black hole binaries based on the observed luminosities \citep{colbert99,kording02,strohmayer09}.

However, the discovery of coherent pulsations in M82 ULX-2 demonstrated unequivocal evidence that the compact object in at least some ULXs is a neutron star \citep{bachetti14}. In fact, it has been proposed that neutron star accretors in ULXs comprise a significant fraction of the ULX population \citep{mushtukov15b,king16,koliopanos17,walton18b}. Since M82 ULX-2, we have discovered ULX pulsars that show persistent pulsations \citep[``persistent" ULX pulsars;][]{furst16,israel17,carpano18,sathyaprakrash19,rodriguezcastillo20} or pulsations that manifest in outbursts \citep[``transient" ULX pulsars;][]{tsygankov17,wilsonhodge18,vasilopoulos20}. The discovery of ULX pulsars have forced a revision in our understanding of the emission mechanisms in these extreme regimes. Proposed mechanisms include the presence of magnetar-like magnetic fields of $B\sim10^{14}$ G \citep{mushtukov15a}, or the beaming of a weak-field ($B\sim10^{11}$ G) pulsar accreting at super-Eddington rates \citep{kluzniak15,king16}. In this work, we considered in particular NGC 300 ULX-1, a high-mass X-ray binary hosting a pulsar \citep{binder11a,heida19}. 

NGC 300 ULX-1 (hereafter called ULX-1) hosts a ULX pulsar originally associated with supernova impostor SN 2010da \citep{binder11a,binder20}. Pulsations with an average value of $31.6$ s were discovered during simultaneous \xmm{} and \nustar{} observations in December 2016 \citep{carpano18}. Monitoring with \chandra, \nicer, \xmm, and \swift{} showed that the spin period evolved from 126 s to 18 s over 4 years \citep{vasilopoulos18}. ULX-1 has also been shown to exhibit timing ``anti-glitches" (i.e., sudden spin-down) during its spin-up due to accretion \citep{ray19}. \cite{vasilopoulos19} has also reported on the long-term spin evolution of ULX-1, and showed evidence that the pulsar had been experiencing a constant mass accretion rate (inferred from roughly constant luminosity) even though the X-ray flux decreased by a factor of $\sim50$. This behavior is similar to that seen in the long-term X-ray flux variations of ULX pulsar NGC 7793 P13 \citep{furst21}. For both ULX-1 and NGC 7793 P13, the authors conjectured that this behavior was due to obscuration of the pulsar possibly due to an outflow \citep{kosec18} or Lense-Thirring precession of the inner accretion disk \citep{middleton18,middleton19,khan22}. This supported the work by \cite{carpano18}, who found a large difference in absorption column density between X-ray observations in 2010 (XMM-Newton) and 2016 (XMM-Newton and NuSTAR), while keeping other spectral parameters tied. Optical observations by the Very Large Telescope/X-shooter have revealed the companion object to be a red supergiant \citep{heida19}. 

In this work, we used data from \nicer, \swift, and \xmm{} to characterize the X-ray spectral evolution of ULX-1. However, a complication in studying ULX-1 with non-imaging detectors is the presence of NGC 300 X-1 (hereafter called X-1), a black hole-Wolf Rayet 32.8 hr eclipsing binary, which is 1.4\arcmin\ away from ULX-1, and is well within the same \nicer{} field-of-view (see Figure \ref{fig:integratedimage}). Thus to account for the emission from the nearby X-1 in order to isolate the ULX-1 emission, we used \swift/XRT measurements to create an orbital timing model with which we applied to dense high signal-to-noise \xmm/EPIC data to characterize the X-1 spectrum. We generated mock spectral data, based on the X-1 spectral parameters, folded through the NICER spectral response. To isolate the spectrum of ULX-1, we subtracted both the modeled component from X-1 and the background spectrum (mostly due to particle interactions with the NICER detectors), determined with the ``3C50" model \citep{remillard22}, from each raw spectrum selected in this investigation. 

In this paper, we presented the spectral evolution of NGC 300 ULX-1 across twelve months of observations, completing the analysis of all available \nicer{} data, by constructing spectra from 5-day averaged data. We took advantage of \nicer{}'s sensitive coverage in soft X-rays to characterize the source as it faded by a factor of $\sim50$ across the observation span. In \S~\ref{sec:observations}, we outline the observations taken for ULX-1 and expound on the data analysis. We present the results in \S~\ref{sec:results}, which will be discussed fully in \S~\ref{sec:discussion}.

\section{Observations and Data Analysis} \label{sec:observations}

\subsection{Swift/XRT Observations} \label{sec:swift}

We made use of \swift/XRT 0.3--10 keV observations in the Photon Counting (PC) readout mode taken over 2016 April 14 to 2019 May 11 with ObsIDs 498340[02-89], 88651001, and 88810002. As detailed below, we were looking for a known periodicity on the order of 32 hr, so the 2.5 s time resolution was sufficient, and we required imaging and spectroscopic capabilities \citep{burrows05}. These data were processed by the standard processing pipeline with the good time interval (GTI) expression criteria $\texttt{ELV}\geq28$, $\texttt{BR\_EARTH}\geq120$, $\texttt{SUN\_ANGLE}\geq45$, $\texttt{ANG\_DIST}\leq0.15$, and $\texttt{MOON\_ANGLE}\geq14$.

	\begin{figure}[t]
		\centering
		\includegraphics[width=\linewidth]{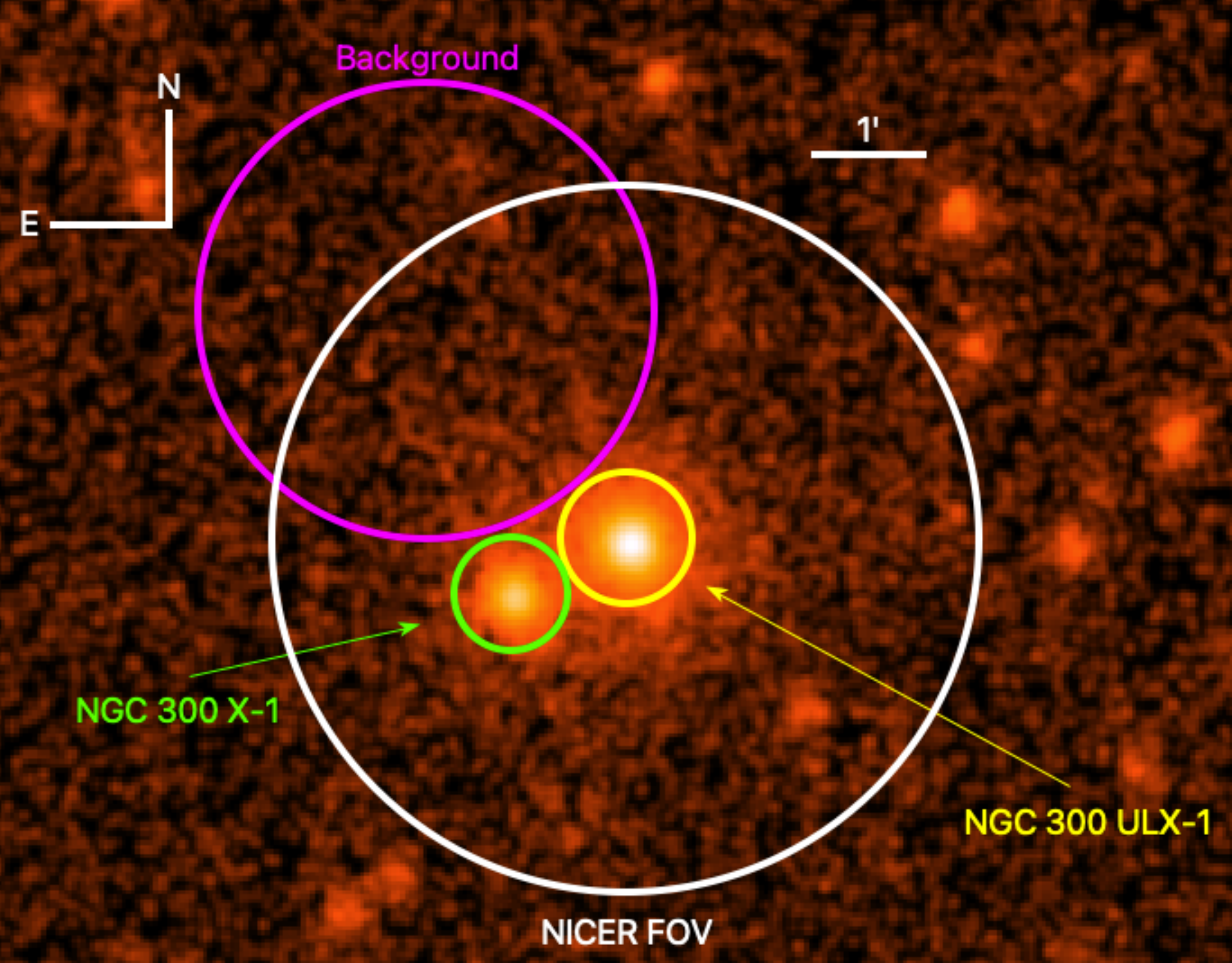}
		\caption{Integrated image of the NGC 300 field from \swift/XRT observations. The image was generated on SAOImageDS9 and smoothed with a Gaussian kernel with radius 3. Events of X-1 (yellow; left) and ULX-1 (blue; right) were taken from extraction regions 30\arcsec\ and 35\arcsec\ in radii, respectively. We used a nearby 120\arcsec\ region for the background of both sources (magenta). The \nicer{} field-of-view (30 arcmin$^2$; white) is also shown for reference, centered on NGC 300 ULX-1. The best known source position for NGC 300 X-1 is $\alpha_{\rm J2000} = 00^{\rm h}55^{\rm m}10\fs00$ and $\delta_{\rm J2000} = -37\degr42\arcmin12\farcs2$ \citep{binder11b}, and for NGC 300 ULX-1 it is $\alpha_{\rm J2000} = 00^{\rm h}55^{\rm m}04\fs85$ and $\delta_{\rm J2000} = -37\degr41\arcmin43\farcs5$ \citep{binder11a}.}
		\label{fig:integratedimage}
	\end{figure}
	
As shown in Figure \ref{fig:integratedimage}, we used 30\arcsec\ and 120\arcsec\ extraction regions for X-1 and the background, respectively, across all the observations; light curves were then extracted with \texttt{XSELECT}, and rate corrections were applied to the light curves with \texttt{xrtlccorr} to account for bad pixels, or point spread function corrections. We also applied barycenter corrections with \texttt{barycorr} with the JPL DE405 solar system ephemeris \citep{standish98}. For the Swift/XRT events, we used \texttt{swxpc0to12s6\_20130101v014.rmf} and \texttt{swxpc0to12s6\_20010101v013.arf} for the redistribution matrix (rmf) and ancillary response (arf) files, respectively.

\subsection{XMM-Newton Observations} \label{sec:xmm}

% Fun fact... SAS is referred to either as Scientific or Science Analysis System. Even the ESA website uses both.

We used \xmm{}/EPIC 0.3--10.0 keV observations of the NGC 300 field which spanned 4 days, from 2016 December 17 to 2016 December 20, with ObsIDs 0791010101 and 0791010301. The \xmm{} Science Analysis System (SAS) v20.0.0 was used for the data reduction, in conjunction with the most up-to-date calibration files as of 2022 June. The EPIC data were reprocessed with \texttt{epproc} and \texttt{emproc} with standard filtering criteria to obtain new event lists. The three EPIC cameras were operated in the imaging data mode (with the medium filter), and we imposed standard filtering criteria where $\textsc{pattern}\leq12$ for MOS cameras, and $\textsc{pattern}\leq4$ and $\textsc{flag} = 0$ for the PN camera. We also filtered the event files for background flares, where we visually inspected the 10.0--12.0 keV light curves to determine the threshold, and applied rate cuts of 0.35, 0.15, and 0.20 c/s for PN, MOS1, and MOS2 cameras, respectively.

The source events for X-1 were extracted using circular regions with radius 25\arcsec\ for all instruments, and the background was extracted from a nearby source-free region with radius 50\arcsec. We also performed barycenter corrections to the cleaned event files with the SAS task \texttt{barycen}, again with the JPL DE405 solar system ephemeris \citep{standish98}. The response files for PN and MOS event files were generated with \texttt{rmfgen} and \texttt{arfgen}.

\subsection{NICER Observations} \label{sec:nicer}

In this subsection, we describe data processing steps that were undertaken with special care to optimize spectral analyses for a source as faint as ULX-1 (0.2--2.0 count s$^{-1}$ for 50 FPMs in the 0.3--4.0 keV range). We also note that NICER is an external payload attached onto the International Space Station (ISS) with 56 co-aligned X-ray concentrators and focal plane modules (FPMs) with silicon drift detectors, of which 52 are operable. The data were processed with version 8 of the NICER Data Analysis Software (\texttt{NICERDAS}), in conjunction with version 6.29 of \texttt{HEASOFT}. We first describe steps to define GTIs, which established the time windows for spectral extractions and background predictions, and then we applied filters to background-subtracted spectra to screen out GTIs that showed signs of inaccuracy in the background prediction. 

We investigated NICER observations of ULX-1 performed from 2018 May 1 through 2019 May 2 (ObsIDs 1034200102 through 2034200205). These data were first brought to a uniform gain calibration by running the HEASARC tool \texttt{nicerl2} on each observation. The calibration file was ``nixtiflightpi20170601v006.fits'', and we note that the last 3 calibrations in this sequence (i.e., v005, v006, and v007) produced only subtle differences, with essentially no changes to the energy conversions in the range suitable for spectral fitting (i.e., 0.3--12.0 keV). The \texttt{nicerl2} task re-runs the data pipeline to determine the time and equivalent photon energy of each recorded event, while also regenerating the ``unfiltered'' and ``cleaned'' event lists for the GTIs found in the reprocessing steps. In the present investigation, we adopted all of the default settings for geometric filters: pointing offset from the target ($<0\fdg015)$, absence in the South Atlantic Anomaly, and minimum angles to the Earth limb and to the solar-illuminated Earth limb ($<15\degr$, and $<30\degr$, respectively). However, the three detector-related filters - maxima for overshoot event rates, undershoot rates, and the relationship between the overshoot rate and the magnetic shielding index - were effectively disabled by specifying impossibly high values for each (i.e., 15000) on the \texttt{nicerl2} command line. The reason for this was to increase the initial exposure time given consideration, with the intent to filter for data quality at a later time, using background-subtracted spectra.

After reprocessing, a list of GTI times was tabulated by running \texttt{nimaketime}, independently, on each observation, using the same combination of filter parameters used previously. However, when defining the final GTIs for spectral extraction purposes, two additional steps were added to facilitate accuracy in the background predictions. We used the ``3C50'' model \citep{remillard22}, an empirical model based on spectral libraries formulated from NICER observations of the designated background fields (RXTE\_BKGD fields). The adjustments to GTI times were intended to limit the dynamic range of 3C50 parameter values, within a GTI, so that non-linearities in the background behavior might be contained when the background model is applied. First, it was found that transitions in ISS day vs. night can reduce the effectiveness of the ``nz'' parameter (raw count rate in 0.0--0.25 keV) for predicting the soft X-ray excess tied to the optical light leak, when such transitions occur within a GTI. In particular, solar reflections off ISS structures in advance of ISS sunrise can cause an early onset of noise increase, while detector noise bleeding after ISS sunset can broaden the noise curve to affect times after a day/night transition. To avoid these effects, we ran \texttt{nimaketime} twice, adding constraints, SUNSHINE=1 and SUNSHINE=0, to each trial, respectively. An initial table of GTI times (MET START and MET STOP) was obtained by converting each GTI file from FITS to ASCII format, while adding a third column to distinguish the SUNSHINE value, and then combining all of the results into one table. GTI sequences can sometimes show brief gaps caused by isolated packet loss or by noise in \texttt{nimaketime} selection parameters, and such gaps can be integrated over without significant consequences. We chose to ignore any gap of 2 s or less, when there was no SUNSHINE transition, and we masked out exposure times $\pm 30$ s from time-adjacent GTIs that are associated with a SUNSHINE transition. 

The other step to improve background predictability is to limit the time range given to each 3C50 model prediction. This is motivated by the fact that NICER background count rate and spectral shape routinely vary in complex ways over the course of the ISS orbit, and also from orbit to orbit. Thus, a single background prediction should not be made for a large interval in ISS orbital phase. On the other hand, the 3C50 model parameters for the stage 1 library (i.e., $ibg$ and $hrej$ - in-focus 15.0--18.0 keV events and rejected particle events near the detector edge, respectively) suffer from low count rates, creating an opposing motivation to integrate as long as possible, to limit the effects of Poisson noise. Noting the four passages between the Earth equator and the highest polar latitudes ($52\degr$) in the ISS orbit (92 min), we chose a target GTI interval of 300 s. Intervals of duration ($dt$) longer than 450 s were subdivided into $N$ GTIs, with $N = int(dt/300 + 0.5)$. The final GTI table was further limited to durations of at least 50 s. The conversion of the initial GTI table to the final one was done with a C program written for this purpose. The final GTI table was indexed, and the index number was included in the filenames of all downstream data products associated with a given GTI. 

For ULX-1, we applied these steps to the selected archive, and this yielded 1057 GTIs and a total exposure of 300.8 ks. This was the starting point for spectral extractions and the application of the 3C50 background model, each conducted per GTI. Extractions of spectra and 3C50 model parameters values were made on the basis of 50 selectable FPMs, while ignoring FPMs 14 and 34, as explained in \cite{remillard22}. For occasions when one or more selectable FPMs were not operating, for a given GTI, then the photon spectra and measurement parameters were linearly rescaled to 50 FPMs, prior to further consideration.

Investigations of NICER spectra should be made with screening efforts to ignore GTIs that have an inaccurate background prediction. Current screening methods use background-subtracted spectra (noted as parameters with subscript ``net''), per GTI. Filtering is performed in diagnostic energy bands that have very low effective area from the NICER optics, leading to the expectation that the net count rate should be near zero. Filtering levels are then advised with criteria depending on source brightness \citep{remillard22}.

In the current investigation, spectral extractions and background modeling were performed for each of the 1057 GTIs. Filtering was then applied at level 3 \citep{remillard22}, selecting $|hbg_{net}| < 0.05$ and $|S0_{net}| < 2.0$, where $hbg_{net}$ is the background-subtracted count rate at 13.0--15.0 keV, and $S0_{net}$ is the analogous rate at 0.2--0.3 keV. This yielded 908 GTIs for ULX-1 with a total exposure time of 263.0 ks, which is 87\% of the total. These were the parent set of GTIs used for the 5-day binning analyses, described in Section \ref{sec:ulx1}.

In this work, we concentrated on the spectral analysis, which was carried out with XSPEC 12.12.1 \citep{arnaud96}. We restricted the energy range to 0.4--4.0 keV due to low source counts in the higher energy band ($>4.0$ keV) and in order to adopt a uniform energy range across all spectra. The spectra were each grouped using the optimal binning scheme and rebinned to have at least 1 count per bin \citep{kaastra16}. The fit statistic employed was the Cash statistic, suitable for the low count rates we are dealing with \citep{cash79,kaastra17}. We also applied a 1\% systematic error to the spectra as recommended by the NICER team\footnote{\url{https://heasarc.gsfc.nasa.gov/docs/nicer/analysis_threads/cal-recommend/}}.

\section{Results} \label{sec:results}

\begin{figure*}[htbp!]
		\centering
		\includegraphics[width=\linewidth]{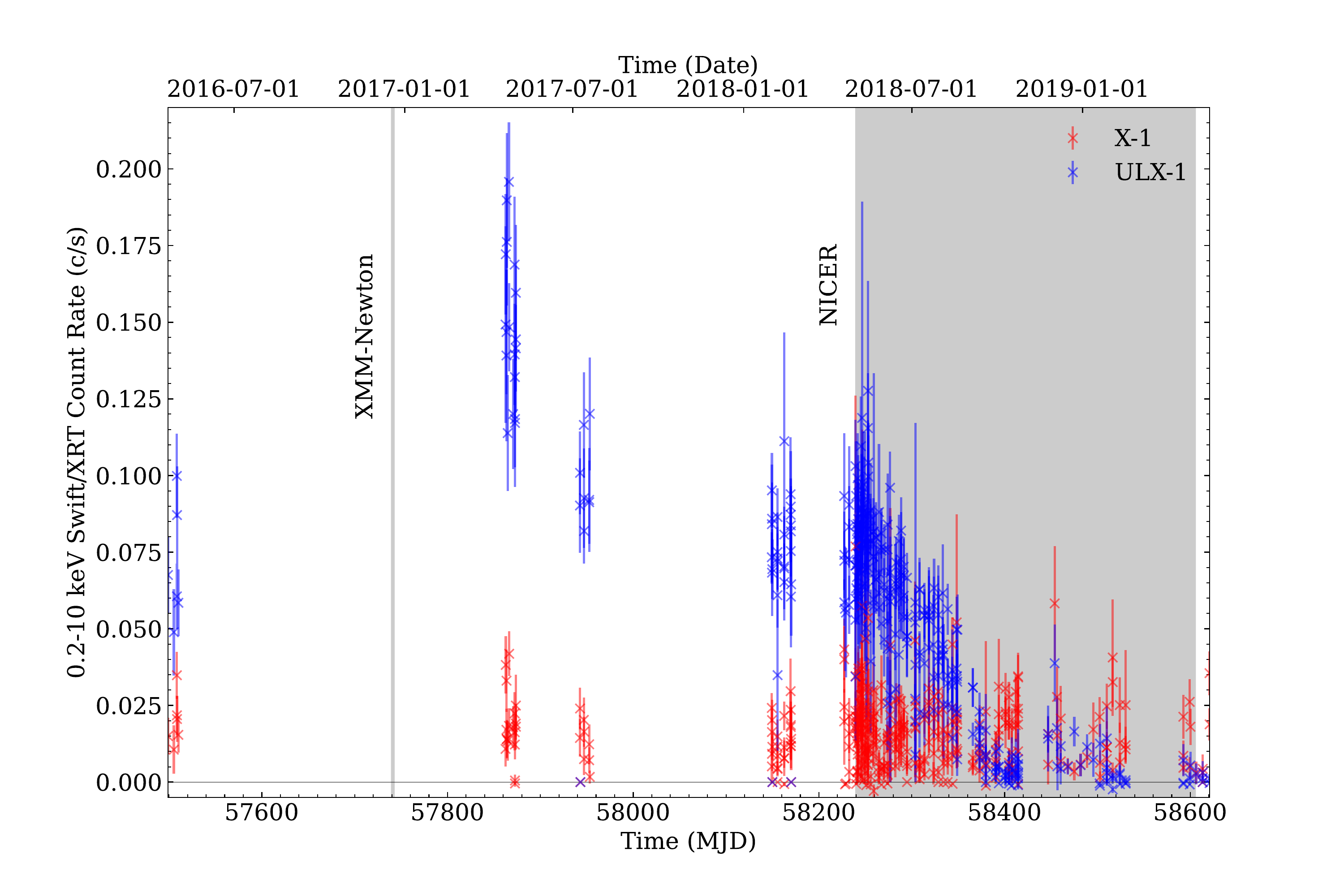}
		\caption{0.2--10.0 keV Swift/XRT background-subtracted and exposure-corrected light curve of NGC 300 X-1 (red) and NGC 300 ULX-1 (blue), binned with 1 hr time bins. The grey shaded regions indicate the time span of the XMM-Newton data (left) and NICER data (right) used in this work.}
		\label{fig:ngc300_lc}
	\end{figure*}

We show the flux history of X-1, and ULX-1 from the Swift/XRT 0.2--10.0 keV events, as well as illustrate the timeline of XMM-Newton and NICER observations, in Figure \ref{fig:ngc300_lc}. X-1 exhibited a constant flux in the long-term, whereas ULX-1 clearly showed a declining flux over the course of the NICER observations. 

\subsection{NGC 300 X-1} 

We attempted to look for variability in the hardness ratio (1.0--2.0 keV/0.4--1.0 keV) that would suggest spectral changes. We constructed 5-day bins with the Swift/XRT data set and did not find any evidence of systematic variability of X-1 within the uncertainties (not plotted). We similarly did not find any systematic trend with the hardness-intensity diagram (with intensity defined over 0.3--10.0 keV).

\begin{figure}[t]
		\centering
		\includegraphics[width=\linewidth]{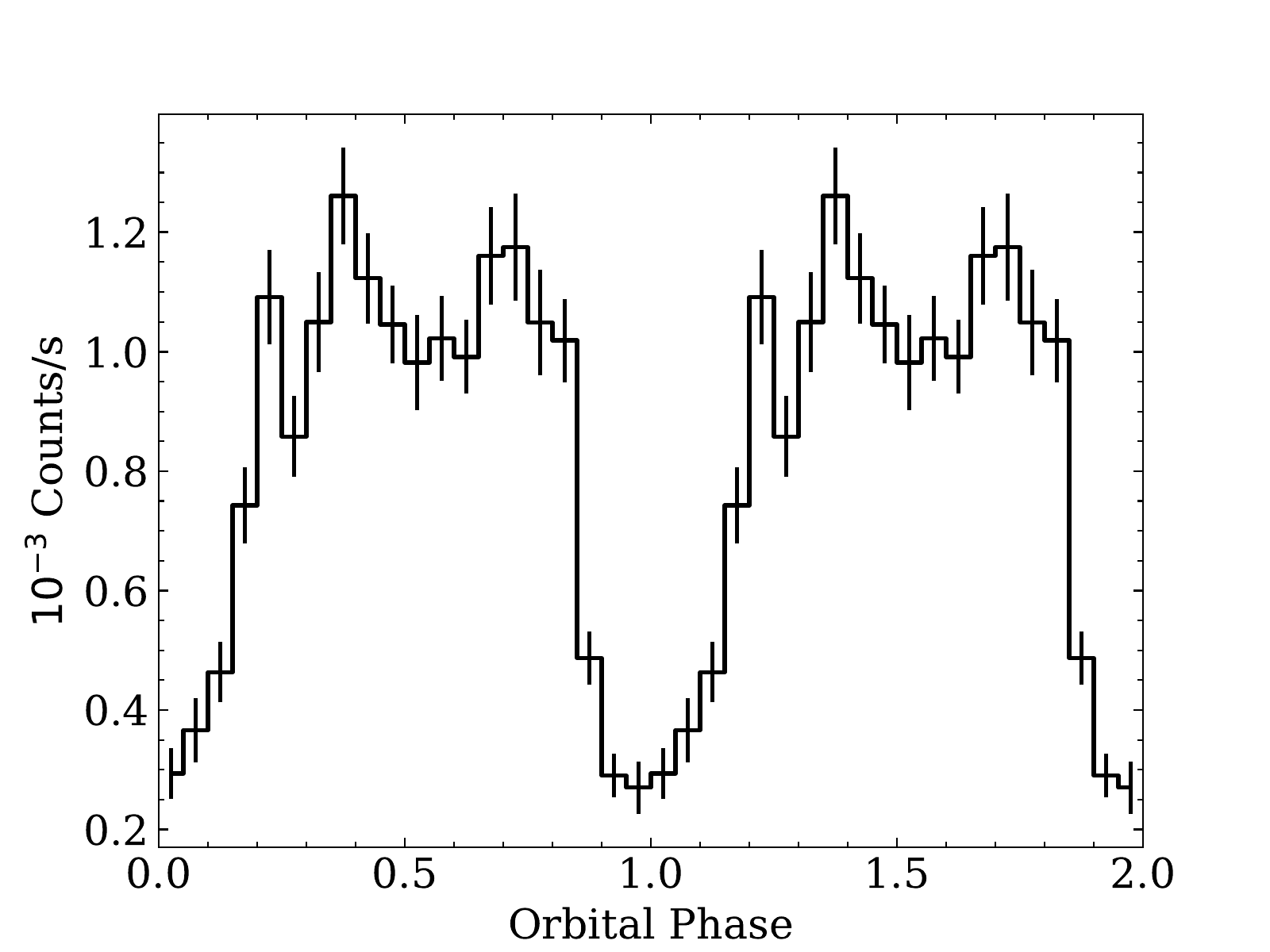}
		\caption{Folded exposure-corrected 0.3--10.0 keV orbital profile with $P_{\rm orb} = 32.788 \pm 0.007$ h and $T_0 = $ MJD 57493.476 $\pm 0.017$ (TDB). The orbital profile has been shifted so that phase 0 corresponds to the eclipse minimum. The orbital profile for the black hole-Wolf Rayet binary NGC 300 X-1 was constructed using events from the Swift/XRT observations that span a time interval that included XMM-Newton and NICER data. Two orbital cycles were plotted for clarity.}
		\label{fig:x1profile}
	\end{figure}

In order to derive an orbital timing ephemeris to construct orbital phase-resolved spectra for X-1, we calculated a Lomb-Scargle periodogram using the Swift/XRT events since we were dealing with an irregularly sampled light curve \citep{lomb76,scargle82,horne86}. First, we noted that \cite{carpano19} combined 86 Swift/XRT PC mode observations over 2006 September 5 to 2018 June 19, 7 XMM-Newton observations over 2000 December to 2016 December, and 5 Chandra observations over 2006 June and 2014 November, to derive an orbital period for X-1 of $32.7932 \pm 0.0029$ h ($1\sigma$). In constructing our ephemeris, we then focused our search around $32.79$ h. To determine the best-fit orbital frequency for our observation, we constructed a grid of trial frequency values and calculated 

\begin{equation}
    \chi^2 = \sum\limits_{i=1}^N \frac{(x_i-\bar{x})^2}{\sigma_i^2},
\end{equation}

where $N$ corresponds to the number of bins in the folded profile, $x_i$ is the count rate in the i-th bin, $\sigma$ is the associated uncertainty, and $\bar{x}$ is the mean count rate across the profile. The uncertainty was calculated from the flux randomization and random subset selection methods \citep{peterson98,welsh99,peterson04}, which accounted for the sampling and measurement uncertainties of the light curves, where we ended up with $\sigma = 0.002$ $\mu$Hz. Thus we found $P_{\rm orb} = 32.788 \pm 0.007$ h, which is formally consistent with the value derived by \cite{carpano19}. The final folded orbital profile is shown in Figure \ref{fig:x1profile}. In order to derive $T_0$, the time of eclipse minimum, we fitted the ramp-and-step model for X-1. The model describes the folded orbital light curve with a constant count rate for the off-eclipse and on-eclipse components, and a linear count-rate evolution during ingress and egress \citep{wachter00,iaria18}. We thus obtained $T_0 = {\rm MJD\,} 57493.476 \pm 0.017$ (TDB) by adding the phase offset due to the eclipse minimum (multiplied by the best-fit orbital period) to the arrival time of the first photon.

With this derived orbital timing ephemeris for X-1, we defined the ``on-eclipse" and ``off-eclipse" segments of X-1, where orbital phases 0.15--0.70 refer to the ``off-eclipse" and orbital phases 0.70--1.15 (modulo 1) refer to the ``on-eclipse", where the black hole is obscured by the dense winds of the companion Wolf-Rayet star. There were too few events to do detailed modeling of the ingress and egress states. We assumed that the spectral behavior did not change significantly throughout the Swift/XRT observation period, which can be inferred from roughly constant spectral hardness in the Swift/XRT spectral data (see first paragraph).

Events for each of these segments were extracted using \texttt{XSELECT} and finally binned into a spectrum, for each of the XMM-PN, XMM-MOS1, and XMM-MOS2 instruments, imposing a grouping with a minimum of 1 count per bin and with the `optimal binning' method \citep{kaastra16}. The ``on-eclipse" and ``off-eclipse" spectra for each of the 3 instruments were simultaneously fit with the C-statistic, i.e., a total of 6 spectra. We found that the model comprising of an absorbed disk blackbody with a Comptonization component and two multiplicative constants best described the spectra, where we employed SIMPL to model the Comptonization component of the plasma surrounding the black hole binary in a simple, self-consistent way \citep{steiner09}. The two multiplicative constants accounted for two different physical effects - the first parameterizes the ratio of the X-1 flux during ``on-eclipse" (set to 1) and ``off-eclipse"; the second parameterized the calibration differences between the three XMM-Newton instruments, with the constant for XMM-PN fixed at 1. Single-component models (with a power law or blackbody) and simple two-component models (absorbed power law and blackbody) inadequately described the X-1 spectra.

We found a reasonable fit with a reduced C-stat of 1.26 (550 d.o.f.), with a photon index of $\Gamma = 2.32_{-0.04}^{+0.04}$, a scattered fraction of $f=0.35$, inner disk temperature of $T_{\rm in} = 0.112_{-0.013}^{+0.017}$ keV, on-eclipse and off-eclipse hydrogen column densities of $n_H = 0.082_{-0.017}^{+0.027} \times 10^{22}{\rm\,cm^{-2}}$ and $n_H = 0.10_{-0.02}^{+0.03} \times 10^{22}{\rm\,cm^{-2}}$, respectively, and a $E = 0.949_{-0.013}^{+0.012}$ keV emission line characterized by a Gaussian with width $\sigma=0.089_{-0.014}^{+0.016}$ keV. The results of the fit are summarized in Table \ref{tab:x1spec}. 

In order to characterize the phase-resolved background contribution from X-1 to use with the NICER data, we used the best-fitting spectral model (see Table \ref{tab:x1spec}) and generated NICER response-folded spectra (using \texttt{fakeit} in \texttt{XSPEC}) that was applied to every spectrum that was extracted for each GTI (defined in \S~\ref{sec:nicer}). We made use of the centroid time of the GTIs to determine the orbital phase and hence whether the GTI was in the on-eclipse or off-eclipse phase, which was valid given the short length of the GTIs (average of 290 s) relative to the much longer orbital period for X-1 ($\sim32.8$ h). When modeling the X-1 spectra as an additional component of "background" for ULX analyses, we assumed no effective area losses for the X-1 spectrum, which was reasonable given that X-1 and ULX-1 are 1.4\arcmin\ apart, and NICER's vignetting profile suggested little attenuation in the response\footnote{https://heasarc.gsfc.nasa.gov/docs/nicer/data\_analysis/ workshops/NICER-CalStatus-Markwardt-2021.pdf} \citep{okajima16}.  

\input{x1spec}

\subsection{NGC 300 ULX-1}\label{sec:ulx1}

\begin{figure}[t]
		\centering
		\includegraphics[width=\linewidth]{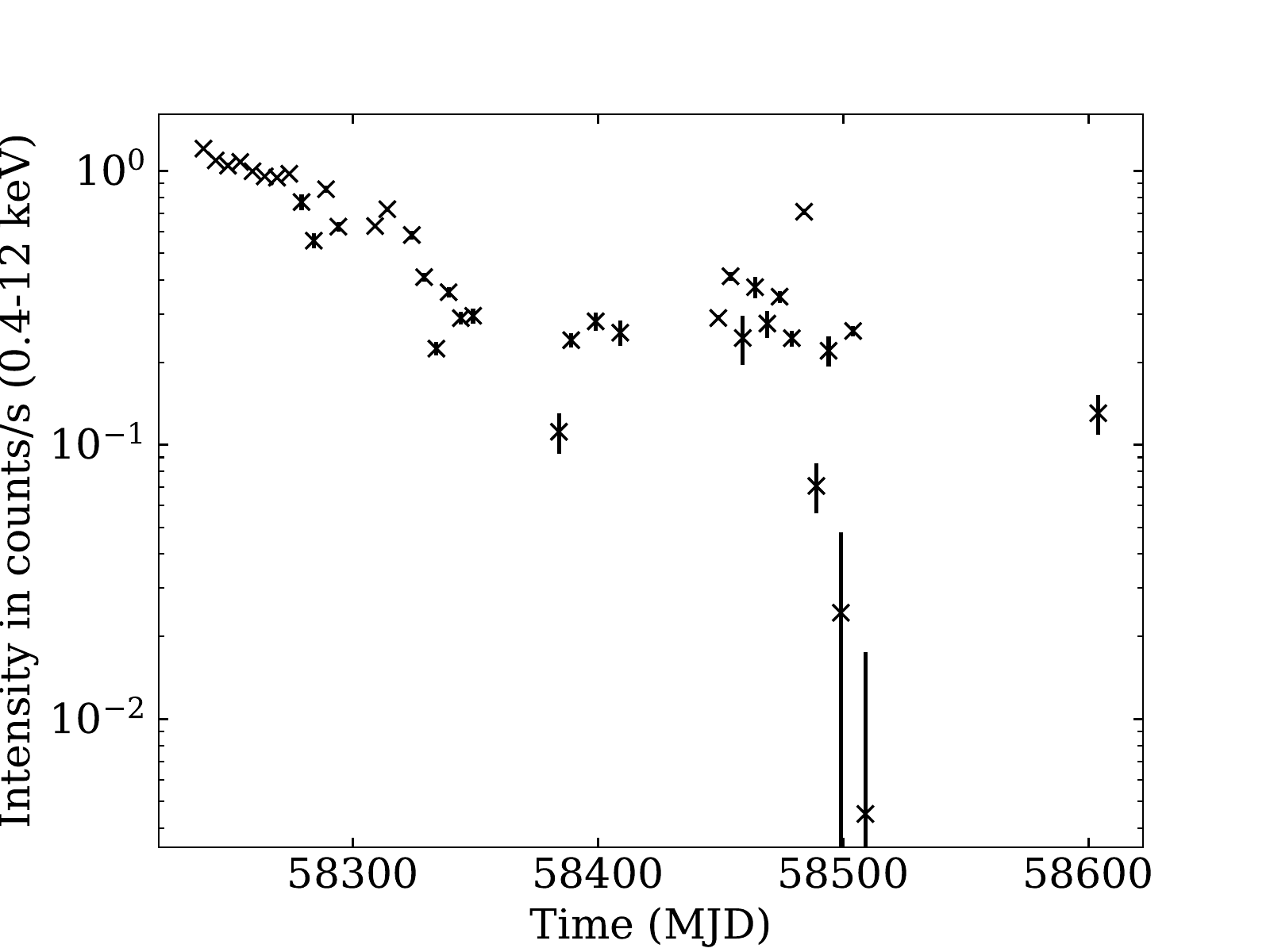}
		\caption{5-day averaged 0.4--12.0 keV light curve of ULX-1 over the span of 12 months of NICER monitoring (2018 May 1 to 2019 May 2), where the particle and X-1 background contributions have been subtracted. The flux from ULX-1 broadly varied by a factor of 25 throughout the span of the NICER observations.}
		\label{fig:ulx_lc}
	\end{figure}

\begin{figure}[t]
		\centering
		\includegraphics[width=\linewidth]{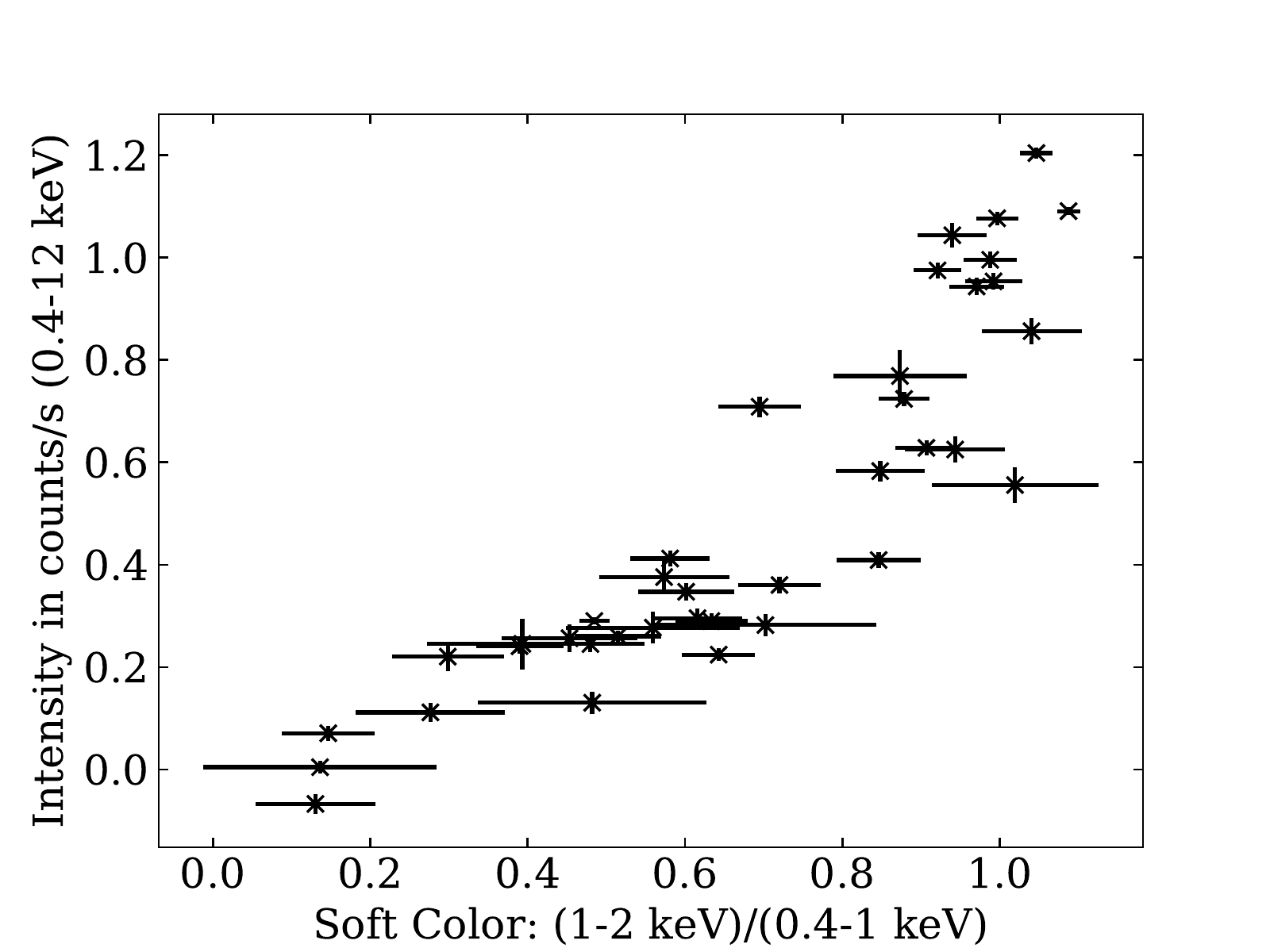}
		\caption{Hardness-intensity diagram of ULX-1 with NICER data where the intensity (0.4--12.0 keV) is plotted as a function of the soft color ((1.0--2.0 keV)/(0.4--1.0 keV)). The intensity is derived using the 5-day averaged spectra. The count rates in each energy band have been corrected for the contributions of the particle and X-1 background (see text for details). There is a clear correlation between the broadband intensity and soft color, where the dimming of the source is accompanied by a steeper spectrum in soft X-rays (i.e., 0.4--2.0 keV).}
		\label{fig:ulx_hid}
	\end{figure}

The nominal 3C50 background predictions for ULX-1 and the NGC 300 X-1 emission model were then combined to create a `total background' measurement for the NGC 300 field to isolate the emission from ULX-1. The spectra from the individual GTIs were then grouped into 5-day bins in order to ascertain the long-term, average behavior of the source and to increase the signal-to-noise. The 0.4--12.0 keV light curve in Figure \ref{fig:ulx_lc} shows a gradual decrease in flux by a factor of $\sim50$ over the one-year span of the data. In Figure \ref{fig:ulx_hid}, we show the hardness-intensity diagram, where we represent spectral hardness with the soft color, defined as the ratio of 1.0--2.0 keV flux to that of 0.4--1.0 keV flux. The hardness-intensity diagram shows a clear dimming of the source as the source softened (i.e., decreasing soft color). 

\begin{figure*}[ht]
		\centering
		\includegraphics[width=\linewidth]{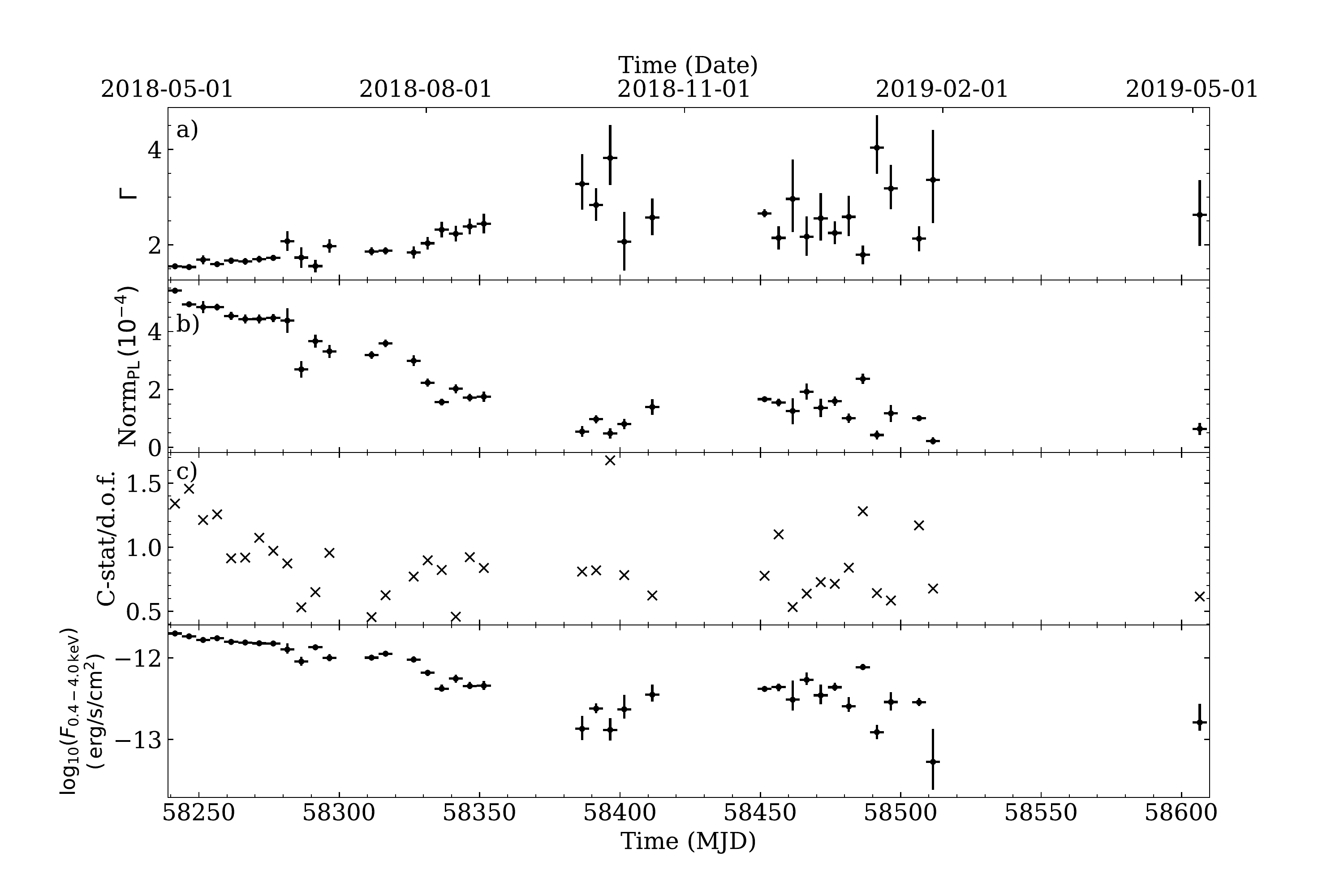}
		\caption{Spectral evolution of NGC 300 ULX-1 with an absorbed power law model as a function of time (MJD), using 1 year of NICER observations with the contribution from X-1 accounted for (details in text). a): Power law index $\Gamma$; b) power law normalization, in units of $10^{-4}$ photons/keV/cm$^2$ at 1 keV; c) ratio of the C-statistic to the number of degrees of freedom (d.o.f.); d) 0.4--4.0 keV absorbed flux in ${\rm erg/s/cm^2}$.}
		\label{fig:specfit}
	\end{figure*}

To understand the long-term evolution of the source, we analyzed the 5-day binned spectra. The source spectra from the individual GTIs in each 5-day bin were simply summed together in counts space with the \texttt{mathpha} tool, whereas the background spectra in the same bin were calculated via a exposure time-weighted average of the individual spectra (i.e., X-1 plus particle background) in rate space also with \texttt{mathpha}. We ended up with 39 spectra, and we performed spectral fits with an absorbed power law and found that it was a good fit overall. We fixed the value of the hydrogen column density at $n_H = 0.11\times10^{22}{\rm\,cm^{-2}}$ \citep{carpano18}. The results of the spectral evolution can be seen in Figure \ref{fig:specfit} with \texttt{tbabs(powerlaw)}, where panel (a) shows the evolution of the power law photon index ($\Gamma$), panel (b) is the power law normalization (in $10^{-4} {\rm\,photons\,keV^{-1}\,cm^{-2}}$), panel (c) shows the reduced C-statistic (C-stat/number of degrees of freedom), and panel (d) plots the 0.4--4.0 keV absorbed flux in ${\rm erg\,s^{-1}\,cm^{-2}}$. We also tried fitting the spectra with absorbed blackbody models (with \texttt{bbodyrad} and \texttt{diskbb}) and found that they were poor fits with reduced C-statistic values above 2. 

\begin{figure}[ht]
		\centering
		\includegraphics[width=\linewidth]{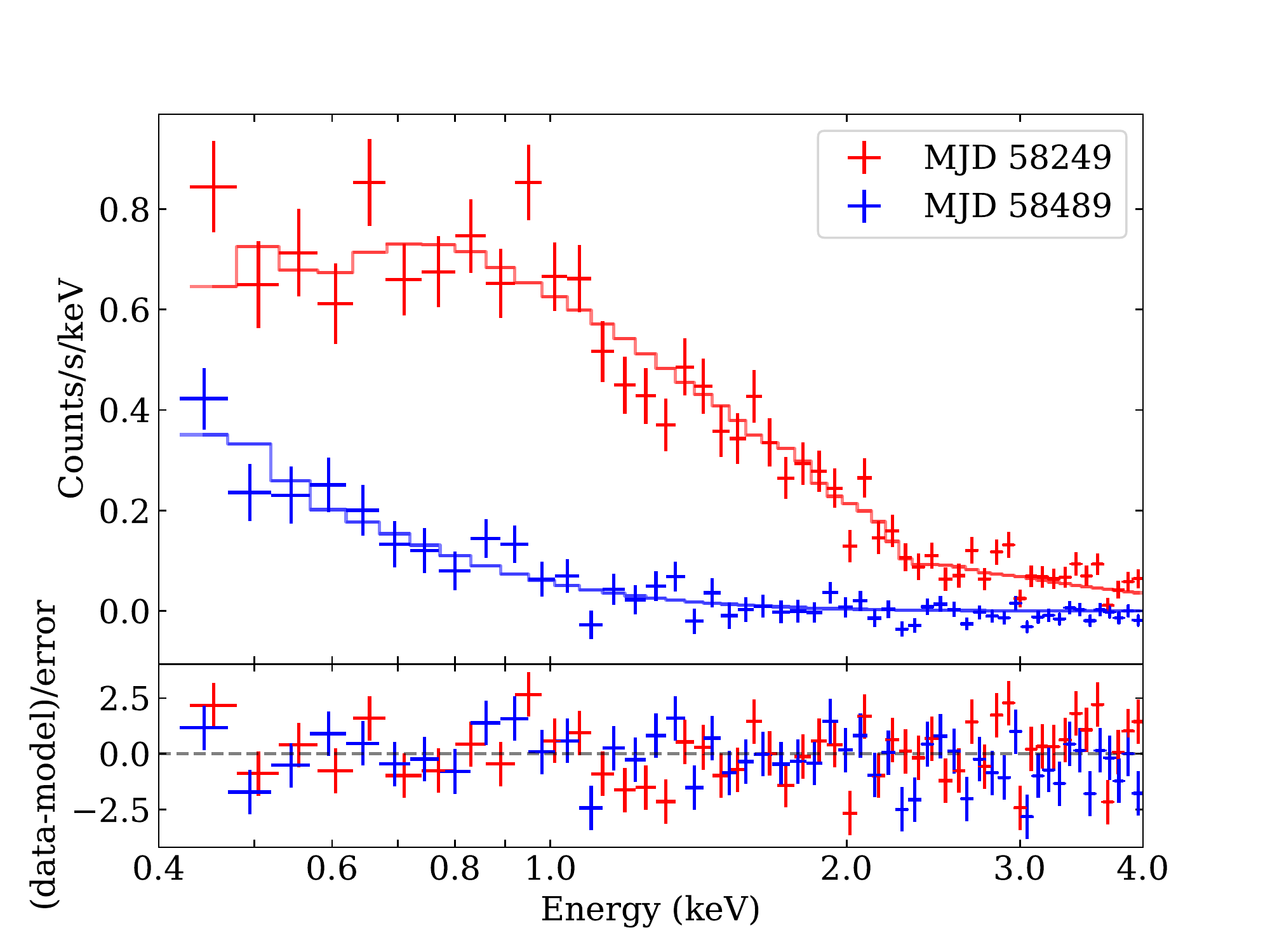}
		\caption{0.4--4.0 keV spectra corresponding to the brighter state of ULX-1 (MJD 58249) and the fainter state (MJD 58489). The absorption column density was fixed at $n_H = 0.11\times10^{22}{\rm\,cm^{-2}}$, and the best-fit power law indices were $\Gamma=1.69_{-0.09}^{+0.09}$ and $\Gamma=4.0_{-0.5}^{+0.7}$ for the brighter and fainter states, respectively.}
		\label{fig:brightfaint}
	\end{figure}

The spectral evolution in Fig. \ref{fig:specfit} matches the hardness-intensity diagram from Fig. \ref{fig:ulx_hid}, where in Fig. \ref{fig:specfit}a, the photon index ($\Gamma$) increased from about $\Gamma\sim1.6$, and reached a maximum value of around $\Gamma\sim4$. The flux meanwhile dropped by a factor of over 50, representing dramatic dimming over the course of the observations. The change in the spectral shape between the bright and faint states of ULX-1 can be seen in Fig. \ref{fig:brightfaint}, where the bright state is shown by the red points (MJD 58249) and the faint state is represented in blue (MJD 58489). 

While an absorbed power law was generally a good fit to the spectra throughout the observation span (see Fig. \ref{fig:specfit}c), some spectra were significantly improved by the introduction of an additional thermal disk blackbody (\texttt{diskbb} in \texttt{XSPEC} parlance) component. We evaluated the improvement of one model relative to the other by considering the difference in the Akaike information criterion \citep[AIC;][]{akaike74}, i.e., $\Delta {\rm AIC}_{12} = {\rm AIC}_1 - {\rm AIC}_2$, where 

\begin{equation}
    {\rm AIC} = -2\text{ln} \mathcal{L} + 2d, 
\end{equation}

where $\mathcal{L}$ is the maximum likelihood of the given model ($\mathcal{L} \propto e^{-C/2}$; $C$ for C-stat), and $d$ is the number of model parameters. We can also reject the competing model at a confidence level by evaluating 

\begin{equation}
    P_1/P_2 \approx e^{-\Delta {\rm AIC}_{12}/2};
\end{equation}

for example, $\Delta {\rm AIC}_{12} = 10$ corresponds to $P_1/P_2 \approx 0.0067$ and is the commonly adopted threshold for decisively distinguishing two models \citep{liddle07,tan12,arcodia18}. The corresponding MJDs, C-statistic values from the fits, as well as the $\Delta {\rm AIC}_{12}$ values, are reported below in Table \ref{tab:aic}. We only reported data points where $\Delta {\rm AIC}_{12}>10$.      

\begin{table*}[t]
\centering
\caption{List of 5-day binned spectra whose fits were significantly improved with an additional thermal disk blackbody model. We only reported the C-statistic values from fits where the inclusion of a disk blackbody component resulted in a $\Delta {\rm AIC}_{12} > 10$.
  \label{tab:aic}
}  
\begin{tabular}{cccccc}
\toprule 
MJD & Temperature & $F_{\rm diskBB}/F_{\rm total}$ & C-stat/d.o.f. & C-stat/d.o.f. & $\Delta {\rm AIC}_{12}$ \\
 & keV & (0.4--4.0 keV)  & (PL only) & (PL + diskBB) & \\
\midrule 
58239 & $0.30_{-0.04}^{+0.04}$ & 0.15 & 75.1/56 & 53.0/54 & 18.2 \\
58244 & $0.34_{-0.03}^{+0.03}$ & 0.20 & 87.4/60 & 45.8/58 & 35.6 \\ 
58254 & $0.27_{-0.04}^{+0.04}$ & 0.16 & 69.1/55 & 46.0/53 & 17.9 \\ 
58454 & $0.19_{-0.05}^{+0.06}$ & 0.27 & 53.9/49 & 29.8/47 & 19.6 \\ 
58484 & $0.19_{-0.05}^{+0.05}$ & 0.20 & 64.0/50 & 27.4/48 & 32.1 \\ 
58504 & $0.21_{-0.04}^{+0.05}$ & 0.30 & 62.0/53 & 34.4/51 & 23.5 \\ 
\bottomrule
\end{tabular} 
\end{table*}

	%\begin{figure}[ht]
	%	\centering
	%	\includegraphics[width=\linewidth]{compare_pl_diskbb.pdf}
	%	\caption{Comparison the spectra from a PL only and a PL+diskbb fit}
	%	\label{fig:pldiskbb}
	%\end{figure}

\section{Discussion} \label{sec:discussion}

We have characterized the 1-year spectral evolution of the dimming ULX pulsar NGC 300 ULX-1 with NICER. Due to the contaminating emission by the nearby (1.4\arcmin) eclipsing X-ray binary, NGC 300 X-1, we used Swift/XRT observations and high-quality XMM-Newton data to generate orbital phase-dependent NGC 300 X-1 (background) spectra to subtract from the raw NICER spectrum of the NGC 300 field. We showed that over time, the softening of ULX-1 accompanied the dimming down to almost undetectable levels in Figs. \ref{fig:ulx_lc} and \ref{fig:specfit}, where the flux dropped by a factor of over 50, and the photon index varied from $\Gamma\approx1.6$ to $\Gamma\approx2.6$ with a significant scatter (see Fig. \ref{fig:specfit}). 

\cite{ray19} investigated the spectral evolution of the pulsed component over an observation epoch that overlapped with the first 40 days of our observation span (MJD 58240 to 58280) and found similar results to this work. However, in their spectral fits, they varied $n_H$ along with $\Gamma$, which introduced a degeneracy in the spectral fitting that made it non-trivial to evaluate the contribution of the power law. Their work also supported our approach of increasing the bin size (to five days) as the source was evolving sufficiently slowly over five days.  

%There have been extensive X-ray spectral characterizations of ULX-1, on both pulsed and pulse phase-averaged X-ray broadband spectra. \cite{walton18a} used the same data set, but analyzed the pulsed spectrum and found that a cutoff power law model coupled with a Comptonization component (scattering seed photons into a power-law component), as well as a cyclotron resonant scattering feature ($\approx$13 keV), provided a good fit. 
%On the other hand, \citep{koliopanos19} adopted a Bayesian analysis framework and concluded that the phase-averaged spectrum is best described with multicolor thermal emission (from an optically thick accretion envelope) with a hard power-law tail. 

The picture that has been emerging for the evolution of ULX-1 is of a pulsar that is being obscured as a result of outflows and/or Lense-Thirring precession causing the plane of the accretion disk to precess along the line-of-sight \citep{carpano18,middleton18,kosec18,vasilopoulos19,binder20}. The Lense-Thirring precession is thought to arise from super-Eddington accretion ($>20 \dot{M}_{\rm Edd}$) building up a large scale-height wind cone outside of the magnetosphere \citep{middleton18,vasilopoulos19,king20}. Other precession mechanisms are also possible, although they are not likely to dominate the Lense-Thirring precession \citep{middleton18,vasilopoulos18,carpano18,walton18a,pan22}. As a result of the obscuration, however, we would also expect to see an increased spectral hardening due to the soft photons being scattered by dust in the accretion disk. On the contrary, we observed a softening of ULX-1 with decreasing intensity over time. We surmise that this could be due to reprocessed emission off the accretion disk that increasingly intrudes on our line-of-sight while the pulsar was obscured. The NICER monitoring campaign allowed us to observe this long term dimming episode as manifested in the observed flux (see Fig. \ref{fig:specfit}d).

We showed representative spectra of the bright (red; MJD 58249) and faint (blue; MJD 58489) states of ULX-1 in Fig. \ref{fig:brightfaint}. During the 2010 outburst from the source, Swift/XRT monitoring showed ULX-1 having a very hard spectrum, with $\Gamma\sim0$ and a 0.3--10.0 keV unabsorbed flux of $8.6\times10^{-13}{\rm\,erg\,s^{-1}\,cm^{-2}}$, five times fainter compared to the brightest state NICER observed around MJD 58239 \citep{binder11a}. Four months after the initial outburst, Chandra observations showed that the flux dropped by a factor of $\sim20$ to $4.5\times10^{-14}{\rm\,erg\,s^{-1}\,cm^{-2}}$ while there was no obvious change in the spectral shape (within uncertainties for the power law index) \citep{binder11a}. Further Chandra observations (fitting over 0.35--8.0 keV) in 2014 spaced six months apart showed a brightening of the source (by a factor of $\sim10$) while there was little evidence of a spectral change, where $\Gamma=0.0\pm0.4$ and $\Gamma=1.2\pm0.8$ in the brighter and fainter states, respectively \citep{binder16}. While we cannot directly compare the photon indices given the difference in the energy range used in the spectral fitting, we also note a similar ``brightening with accompanying spectral hardening" behavior in the NICER data (see Fig. \ref{fig:ulx_hid}. As mentioned above, \cite{carpano18} attributed the spectral differences to a large difference in absorption column density between 2010 (XMM-Newton) and 2016 (XMM-Newton and NuSTAR) observations of ULX-1.

Our findings provided qualitative confirmation of the common interpretation of the source evolution, where previous work by \cite{carpano18} and \cite{vasilopoulos19} have shown that the decreasing flux of ULX-1 can be explained by increasing obscuration along the line-of-sight. However, in our work, we opted to fix the interstellar line-of-sight absorption at $0.11\times10^{22}{\rm\,cm^{-2}}$ (as in \cite{carpano18}), and used the photon index ($\Gamma$) as our proxy to characterize the ULX-1 spectral evolution. While this is not a physical model of the source, given the low number of counts in the NICER observations, we would not be able to distinguish between simple spectral models (e.g., absorbed power law) and more complex spectral models that include variable absorption and a scattering component (model parameter degeneracy). We also remark that in our work, we are reporting results from a one-year monitoring campaign of ULX-1 with amply dense coverage. While the energy range we employed for the spectral fitting is narrow (0.4--4.0 keV), we are focusing on the long-term spectral evolution of the source rather than diagnosing the source state at any given time.

While we did not have sufficient statistics to compare the partial covering absorber models to the absorbed power law model in a robust manner like in \cite{carpano18}, we explored the possibility of short-term variability of ULX-1 that could be explained by a partial covering absorber. 
%As mentioned in \S \ref{sec:results}, we have attempted to fit single-component absorbed blackbody models to the 39 spectra and found very poor fits overall. However, we lacked the signal-to-noise necessary to completely exclude these models for some spectra. In spite of that, 
There were some 5-day binned spectra whose fits were significantly improved by including an additional thermal disk blackbody component. The significance was assessed using differences in the AIC values between the absorbed power law and the absorbed power law and disk blackbody models (see Table \ref{tab:aic}); we surmise that this provided some suggestive evidence of the `appearance' of the pulsar in between obscuration episodes by partial covering absorbers, thus opening up the detection of X-ray pulsations and thermal emission. 

It is intriguing to note from Table \ref{tab:aic} that significant fit improvements from the addition of a disk blackbody traced the detection of X-ray pulsations somewhat -- the additional thermal component is well-detected in some time intervals prior to MJD 58349, and then pulsations disappeared until MJD 58454, around which pulsations were detected once again until around MJD 58484 \citep{vasilopoulos19}. Thus we suggest a link between the appearance of a disk blackbody component and the presence of X-ray pulsations. \cite{ray19} also looked at the long-term spin frequency evolution of the ULX pulsar with the data at MJD 58155, as well as between MJD 58239 and MJD 58411; they found that the pulsar was spinning up, and stopped detecting significant pulsations after around MJD 58350. Intriguingly, we found a significant thermal component in the ULX-1 spectrum around MJD 58504, but no pulsations seemed to have been detected around that epoch down to a significance of $4.5\sigma$ \citep{vasilopoulos19}.

Optical spectroscopy has identified the companion to be a red supergiant (RSG), which suggested that the orbital period of the ULX-1 system was at least 0.8--2.1 yr \citep{heida19}. It is noted that NGC 7793 P13 is the only other known ULX pulsar with a supergiant companion (spectral classification B9Ia) \citep{motch11,motch14}. \cite{townsend20} derived an empirical relationship between the orbital period ($P_{\rm orb}$) and superorbital period ($P_{\rm sup}$) using a sample of high mass X-ray binaries and pulsar ULX systems, where $P_{\rm sup} = (22.9\pm0.1)P_{\rm orb}$. If we assume the orbital period for ULX-1 to be 0.8--2.1 yr, then the superorbital period (i.e., precession period) could be at least 18--41 years, and could be even higher for larger orbital periods \citep{heida19,townsend20}. This means that it could be decades before sustained pulsation detections from the ULX pulsar will come into full view again; and we are currently observing a persistent low-lying flux level punctuated by occasional detections of the source (and pulsations) due to the partial covering absorber (as mentioned through Table \ref{tab:aic}). 

The observed behavior from ULX-1 suggests that unlike several known transient ULX pulsars \citep{tsygankov17,wilsonhodge18,vasilopoulos20} that show pulsations in outbursts and variable mass accretion rates, ULX-1 is more like the persistent ULXs \citep{bachetti14,furst16,israel17,sathyaprakrash19,rodriguezcastillo20} given the long-term behavior we see, where the mass accretion rate is constant \citep{vasilopoulos19}. This behavior is also similar to that seen in NGC 7793 P13 \citep{furst21}, where long-term monitoring (spanning about four years) of the source evolution showed that while the source X-ray flux was dimming, the period derivative was not correlated with the X-ray flux, suggesting that a variable accretion rate was not responsible for the dimming X-ray flux.

Finally, we remark that in the spectrum for the ``off-eclipse" in NGC 300 X-1, we found a Gaussian emission line with centroid energy $E = 0.949_{-0.013}^{+0.012}$ keV and width $\sigma = 0.089_{-0.014}^{+0.016}$ keV, which could be associated with Fe-L emission or with the Ne X line \citep{degenaar13,bult21}. It could also be due to Fe XVIII from photoionized material originating in the outer edge of the accretion disk, as seen in Cyg X-2 \citep{vrtilek86,chiappetti90}, or the complex blending of Fe and O emission lines \citep{vrtilek88}.

%Farinelli+ 09 had an exposition on the 1 keV line. 

\acknowledgments

We thank the anonymous referee for a careful review of the manuscript. We also thank Georgios Vasilopoulos for initial discussions that led to refining this work. M.N., R.R., D.P., and this work were supported by NASA under grant 80NSSC19K1287. This research has made use of data and/or software provided by the High Energy Astrophysics Science Archive Research  Center (HEASARC), which is a service of the Astrophysics Science Division at NASA/GSFC and the High Energy Astrophysics Division of the Smithsonian Astrophysical Observatory.

%Was I also technically paid by 80NSSC19K0634? 

\facilities{NICER, XMM, Swift}

\software{Astropy \citep{astropy:2013, astropy:2018}, NumPy and SciPy \citep{virtanen20}, Matplotlib \citep{hunter07}, IPython \citep{perez07}, tqdm \citep{dacostaluis22}, HEASoft 6.29\footnote{http://heasarc.gsfc.nasa.gov/ftools} \citep{heasoft}}

%\appendix

%\section{Appendix information}

\bibliography{ulx}{}
\bibliographystyle{aasjournal}

\end{document}

%% file: x1spec.tex
%total_varytbnew(simpl(diskbb+gauss))-const-const-3.xcm (see note from 09/22/2020 and 09/23/2020)
\begin{table}[t]
\centering
\caption{Best-fit spectral parameters from simultaneously fitting 0.3--10.0 keV XMM-Newton spectra for the ``off-eclipse" and ``on-eclipse" components of NGC 300 X-1. The spectra are described with a phenomenological model with the model \texttt{tbnew(simpl(diskbb+gaussian))*constant} in \texttt{XSPEC} parlance.
  \label{tab:x1spec}
}  
\begin{tabular}{lcc}
\toprule 
Parameter & Value & Units \\
\midrule 
\underline{Tied} & \\
Photon Index & $2.32_{-0.04}^{+0.04}$ & \nodata \\
Scattered Fraction & $0.35_{-0.10}^{+0.10}$ & \nodata \\
$T_{\rm in}$ & $0.112_{-0.013}^{+0.017}$ & keV \\
norm (diskbb) & $104_{-61}^{+151}$ & \red{\nodata} \\
\midrule 
\underline{Constants} & \\ 
Flux & $0.50_{-0.02}^{+0.03}$ & \nodata \\
MOS1 & $1.08_{-0.03}^{+0.03}$ & \nodata \\
MOS2 & $1.14_{-0.03}^{+0.03}$ & \nodata \\ 
\midrule 
\underline{On-eclipse} & \\
$n_H$ & $0.082_{-0.017}^{+0.027}$ & $10^{22}{\rm\,cm^{-2}}$ \\ 
\midrule
\underline{Off-eclipse} & \\ 
$n_H$ & $0.10_{-0.02}^{+0.03}$ & $10^{22}{\rm\,cm^{-2}}$ \\
LineE (keV) & $0.949_{-0.013}^{+0.012}$ & keV \\
Sigma (keV) & $0.089_{-0.014}^{+0.016}$ & keV \\ 
norm (Gauss) & $1.40_{-0.18}^{+0.21}\times10^{-5}$ & photons s$^{-1}$ cm$^2$ \\
\midrule 
C-stat/d.o.f. & 693.92/550 & \nodata \\ 
\bottomrule
\end{tabular} 
\end{table}